\documentclass[aps,pre,amsmath,amssymb,twocolumn,groupedaddress]{revtex4-1}

\usepackage{graphicx}

\usepackage{algorithm}
\usepackage{algpseudocode}
\usepackage{textcomp}

\newcommand{\mean}[1]{\ensuremath{\left<#1\right>}}
\newcommand{\meanalt}[1]{\ensuremath{\langle#1\rangle}}

\newcommand{\df}{\ensuremath{\mathrm{d}}}
\newcommand{\ddt}{\ensuremath{\frac{\df}{\df t}}}

\newcommand{\vnu}{\ensuremath{\nu}}
\newcommand{\vth}{\ensuremath{\theta}}
\newcommand{\vTh}{\ensuremath{\Theta}}
\newcommand{\vecr}{\ensuremath{r}}
\newcommand{\vs}{\ensuremath{s}}

\newcommand{\pE}{\ensuremath{\mathbf{E}}}
\newcommand{\pP}{\ensuremath{\Pr}}

\newcommand{\opF}{\ensuremath{\mathcal{F}}}
\newcommand{\opL}{\ensuremath{\mathcal{L}}}
\newcommand{\opP}{\ensuremath{\mathcal{P}}}
\newcommand{\opQ}{\ensuremath{\mathcal{Q}}}
\newcommand{\opS}{\ensuremath{\mathcal{S}}}

\newcommand{\natz}{\ensuremath{\mathbb{N}_0}}

\newcommand{\vxs}{\ensuremath{x}}
\newcommand{\vxe}{\ensuremath{\env{x}}}
\newcommand{\vXs}{\ensuremath{X}}
\newcommand{\vXe}{\ensuremath{\env{X}}}
\newcommand{\vms}{\ensuremath{y}}

\newcommand{\vMs}{\ensuremath{Y}}
\newcommand{\vMe}{\ensuremath{\env{Y}}}

\newcommand{\spaceS}{\ensuremath{\natz^{N}}}
\newcommand{\spaceE}{\ensuremath{\natz^{\env{N}}}}
\newcommand{\spS}{\ensuremath{\mathbb{X}}}
\newcommand{\spE}{\ensuremath{\env{\mathbb{X}}}}

\newcommand{\Ns}{\ensuremath{N}}
\newcommand{\Ne}{\ensuremath{\env{N}}}

\newcommand{\JZ}{\ensuremath{\overline{J_{\mathbb{X}}}}}
\newcommand{\notJZ}{\ensuremath{J_{\mathbb{X}}}}

\newcommand{\env}[1]{\ensuremath{\hat{#1}}}
\newcommand{\species}[1]{\ensuremath{\textrm{#1}}}
\newcommand{\adj}[1]{\ensuremath{#1^{\dagger}}}

\begin{document}
\noindent\hspace{-270pt}{\textcopyright 2018 American Physical Society.}

\title{Marginal process framework:\\A model reduction tool for Markov jump processes}

\author{Leo Bronstein}

\author{Heinz Koeppl}
\email[]{heinz.koeppl@bcs.tu-darmstadt.de}

\affiliation{Department of Electrical Engineering and Information Technology, Technische Universit\"at Darmstadt, 64283 Darmstadt, Germany}


\begin{abstract}
Markov jump process models have many applications across science. Often, these models are defined on a state-space of product form and only one of the components of the process is of direct interest. In this paper, we extend the marginal process framework, which provides a marginal description of the component of interest, to the case of fully coupled processes. We use entropic matching to obtain a finite-dimensional approximation of the filtering equation, which governs the transition rates of the marginal process. The resulting equations can be seen as a combination of two projection operations applied to the full master equation, so that we obtain a principled model reduction framework. We demonstrate the resulting reduced description on the totally asymmetric exclusion process. An important class of Markov jump processes are stochastic reaction networks, which have applications in chemical and biomolecular kinetics, ecological models and models of social networks. We obtain a particularly simple instantiation of the marginal process framework for mass-action systems by using product-Poisson distributions for the approximate solution of the filtering equation. We investigate the resulting approximate marginal process analytically and numerically.
\end{abstract}

\pacs{}

\maketitle

\section{Introduction}
Markov jump processes (MJP) have many applications across science and engineering.
The master equation (ME) \cite{vankampen1983}, which governs the time evolution of the probability distribution of the process, is generally too complicated to solve analytically, and frequently infeasible to solve numerically.
A popular alternative is the stochastic simulation (Gillespie) algorithm \cite{gillespie1977}, which produces samples of trajectories of the process.
For larger systems, however, this approach can be computationally very expensive, especially if the system exhibits multi-scale behavior.

In many cases, the MJP is defined on a product-form state-space $\spS \times \spE$ and only one component is of direct interest, while the other component can be considered a nuisance variable.
The question arises whether it is possible to derive a reduced description for the stochastic process corresponding to the component of interest only.
Mathematically, the remaining component then has to be marginalized out of the full stochastic process.
This approach has been used for reaction networks in the case when the variables of interest do not influence the nuisance variables \cite{zechner2014}, and also in other contexts \cite{giesecke2011}, mostly to speed up stochastic simulations.

A large number of other model reduction methods have been proposed in the literature. Many of these are based on time-scale separation or abundance separation, a non-exhaustive list being \cite{cao2005, hellander2007, jahnke2011, thomas2012, constable2013, kang2013, kim2017}.
Other approaches based on marginalization have recently been published \cite{rubin2014, bravi2016, bravi2017} and focus on stochastic differential equation models.

In this article, we extend the marginalization approach of \cite{giesecke2011, zechner2014} to a general MJP with full coupling between variable of interest and nuisance variable.
Marginalization requires the solution of the (in general, infinite-dimensional) filtering equation, which describes the evolution of the conditional probability of the nuisance variable given the trajectory of the marginal process.
We use entropic matching \cite{ramalho2013, bronstein2017} to obtain a finite-dimensional approximation. The filtering equation and entropic matching can be interpreted as the result of projection operations consecutively applied to the full ME of the joint process. In this way, we obtain a principled model reduction method.
Our focus is on the marginal process framework as a theoretical tool for model reduction, rather than as a method for more efficient stochastic simulation.

A particularly important class of MJPs with applications in chemical kinetics, biological population models and models of social interactions are reaction networks.
For reaction networks with mass-action kinetics, a particularly simple reduced description is obtained when a product-Poisson ansatz distribution is used for the approximate solution of the filtering equation. We call the resulting reduced model the \emph{Poisson-marginal process} and investigate it in detail.
Analogously, for exclusion processes, a product-Bernoulli ansatz distribution leads to what is the simplest possible reduced model within our framework. We investigate this reduced process for the example of the totally asymmetric exclusion process (TASEP) on the line with open boundaries.

This paper is organized as follows:
After describing the problem setting in Sec.~\ref{sec:setting}, we provide an outline of the proposed method in Sec.~\ref{sec:outline}, using a simple model of constitutive gene expression as a running example.
The general form of the marginal process framework is derived in Sec.~\ref{sec:marginal}.
The finite-dimensional approximations necessary for a tractable description of the marginal process are discussed in Sec.~\ref{sec:finite_dim}, where we also apply our method to the TASEP as a first example. The Poisson-marginal process for mass-action reaction networks is discussed in Sec.~\ref{sec:poisson}.


\section{Setting} \label{sec:setting}
We consider an MJP $(X, \env{X}) = (X(t), \env{X}(t))_{t \geq 0}$ on a product-form state-space $\spS \times \spE$, where $\spS$ and $\spE$ are countable sets.
The marginal probability distribution $p_t(x, \env{x}) = \pP(X(t) = x, \env{X}(t) = \env{x})$ of such a process (with initial distribution $p_0(x, \env{x})$ at time 0) is governed by the ME
\begin{align*}
&\ddt p_t(x, \env{x})
= [\opL p_t](x, \env{x}) \\
&= \sum_{y, \env{y}} \big\{ L(x, \env{x} \mid y, \env{y}) p_t(y, \env{y}) - L(y, \env{y} \mid x, \env{x}) p_t(x, \env{x}) \big\},
\end{align*}
where $L(x, \env{x} \mid y, \env{y})$ is the rate of transitioning from state $(y, \env{y})$ to state $(x, \env{x})$.
We will also require the backwards evolution operator $\opL^{\dagger}$, which acts on functions $\psi: \spS \times \spE \rightarrow \mathbb{R}$ and is given by
\begin{equation} \label{eq:mjp_cme_backward}
[\adj{\opL} \psi](x, \env{x}) 
= \sum_{y, \env{y}} L(y, \env{y} \mid x, \env{x}) \left\{ \psi(y, \env{y}) - \psi(x, \env{x})\right \}.
\end{equation}
It is the adjoint of $\opL$ with respect to the pairing $(p, \psi) := \sum_{x, \env{x}} p(x, \env{x}) \psi(x, \env{x})$.
Recall that $\opL^{\dagger}$ governs the moment equations of the stochastic process. Thus, for a function $\psi(x, \env{x})$, the expectation $\mean{\psi}_t$ with respect to the distribution $p_t(x, \env{x})$ evolves according to
\begin{equation} \label{eq:derivation_moment_eq}
\begin{aligned}
\ddt \mean{\psi}_t &= \sum_{x, \env{x}}{ \psi(x, \env{x}) \ddt p_t(x, \env{x}) }
= \sum_{x, \env{x}}{ \psi(x, \env{x}) [\opL p_t](x, \env{x}) }\\
&= \sum_{x, \env{x}}{ p_t(x, \env{x}) [\opL^{\dagger} \psi](x, \env{x}) }
= \mean{\opL^{\dagger} \psi}_t.
\end{aligned}
\end{equation}
We are particularly interested in reaction networks, consisting of $N + \env{N}$ species and $R$ reactions that are specified as
\begin{equation} \label{eq:reaction_system}
\sum_{n=1}^{N}{s_{nj} \species{X}_n} + \sum_{n=1}^{\env{N}}{\env{s}_{nj} \env{\species{X}}_n} \longrightarrow \sum_{n=1}^N{r_{nj} \species{X}_n} + \sum_{n=1}^{\env{N}}{\env{r}_{nj} \env{\species{X}}_n}
\end{equation}
for $j = 1, \dots, R$. Here we have divided the set of all species into the species $\species{X}_1, \dots, \species{X}_{N}$ of interest, the \emph{subnet}, and the remaining species $\env{\species{X}}_1, \dots, \env{\species{X}}_{\env{N}}$, the \emph{environment}.
We are interested in the case when the copy numbers of some of the species might be small, so that a fully stochastic description in terms of an MJP is necessary.
The process $X = (X_1, \dots, X_N)$ then describes the state of the subnet species, while $\env{X} = (\env{X}_1, \dots, \env{X}_{\Ne})$ describes the state of the environment species.
The state-space is given by $\spS = \spaceS$ and $\spE = \spaceE$.
For a state $(x, \env{x})$, for each $j = 1, \dots, R$ there exists a transition to the state $(x + \nu_j, \env{x} + \env{\nu}_j)$ with rate $h_j(x, \env{x})$ and change vector $(\nu_j, \env{\nu}_j)$ with $\nu_j = (r_{1j}-s_{1j}, \dots, r_{Nj}-s_{Nj})$ and $\env{\nu}_j = (\env{r}_{1j}-\env{s}_{1j}, \dots, \env{r}_{\Ne j}-\env{s}_{\Ne j})$.
The operators $\opL$ and $\adj{\opL}$ take the form
\[
\begin{aligned} \label{eq:reaction_network_operators}
[\opL p](\vxs, \vxe)
&=\sum_{j=1}^R\bigg\{ h_j(\vxs-\vnu_j, \vxe-\env{\vnu}_j)p(\vxs-\vnu_j, \vxe-\env{\vnu}_j) \\
&\quad \quad \quad \;\,  -h_j(\vxs, \vxe)p(\vxs, \vxe)\bigg\},\\
[\opL^{\dagger} \psi](\vxs, \vxe) 
&= \sum_{j=1}^R h_j(\vxs, \vxe)\left\{\psi(\vxs + \nu_j, \vxe + \env{\nu}_j) - \psi(\vxs, \vxe)\right\}.
\end{aligned}
\]
In all examples treated in this article, we will employ mass-action kinetics, which are given by $h_j(\vxs, \vxe) = \Omega c_j f_j(\vxs) \env{f}_j(\vxe)$ with
\begin{equation} \label{eq:mass_action}
f_j(\vxs)  = \prod_{n=1}^{\Ns}{\frac{(x_n)_{s_{nj}}}{\Omega^{s_{nj}}}},
\quad \env{f}_j(\vxe) = \prod_{n=1}^{\Ne}{\frac{(\env{x}_n)_{\env{s}_{nj}}}{\Omega^{\env{s}_{nj}}}},
\end{equation}
where $(x)_s = x(x-1)\cdots(x-s+1)$ denotes the falling factorial, $c_j$ is a reaction rate constant and where we have introduced the system size $\Omega$ in terms of which we will analyze the behavior of the approximate marginal process.

We will also require a description of the process $(\vXs, \vXe)$ in terms of the $R$ reactions. We associate with each reaction channel $j$ a counting process $Y_j(t)$ which counts the number of firings of reaction $j$ over the time interval $[0, t]$.
The process $(Y_1, \dots, Y_R)$ can again be seen as a reaction network of the form \eqref{eq:reaction_system} with values in $\natz^R$, which can only change by increments of size 1 in any one of its components at a single time.
The state of the original process $(\vXs, \vXe)$ is recovered from the state of these counting processes via
\[
X(t) = \sum_{j=1}^R Y_j(t) \vnu_j, \quad \env{X}(t) = \sum_{j=1}^R Y_j(t) \env{\vnu}_j.
\]

\section{Outline of the method} \label{sec:outline}
As explained in the introduction, our goal is to derive a marginal process description for the process of interest $X$.
While the joint process $(X, \env{X})$ is Markovian, this is no longer the case for the marginal process $X$. The effect of the nuisance variables $\env{X}$ is implicitly contained in the memory of the process $X$.
We now illustrate our proposed method on a simple reaction network from biology. 
We focus on the underlying ideas and postpone derivations to later sections.

A very simple model of constitutive gene expression is given by the reaction network
\begin{equation} \label{eq:network_geneexp}
\begin{aligned}
\emptyset &\overset{c_1}{\longrightarrow} \species{mRNA} \overset{c_2}{\longrightarrow} \emptyset,\\
\species{mRNA} &\overset{c_3}{\longrightarrow} \species{mRNA} + \species{Protein},\quad \species{Protein} \overset{c_4}{\longrightarrow} \emptyset.
\end{aligned}
\end{equation}
Assuming that we are interested primarily in the protein dynamics, we will consider the mRNA to be a nuisance species. Our goal is to obtain a marginal description of the protein dynamics.
Thus, the mRNA plays the role of the nuisance variable $\env{X}$ and the protein the role of the variable of interest $X$.
Assuming mass-action kinetics, the transition rates of the four reactions in the state $(x, \env{x})$ are given in Table~\ref{table}.
\begin{table}
\begin{ruledtabular} 
\begin{tabular}{cccc}
$\emptyset \rightarrow \species{mRNA}$ & $\species{mRNA} \rightarrow \emptyset$
& $\species{mRNA} \rightarrow \species{mRNA} + \species{P}$ & $\species{P} \rightarrow \emptyset$ \\
\hline
$\Omega c_1$ & $c_2 \env{x}$ & $c_3 \env{x}$ & $c_4 x$ \\
\end{tabular}
\end{ruledtabular}
\caption{Transition rates for the reaction network \eqref{eq:network_geneexp} when the process is in state $(x, \env{x})$. \label{table}} 
\end{table}

The steps to obtain a tractable approximate description of the marginal process are as follows:
(i) Determine how the transition rates of the marginal process at time $t$ depend on the process history $x_{[0,t]}$.
(ii) Find a description for these marginal transition rates in terms of an evolution equation driven by the process history $x_{[0,t]}$. The resulting equations are generally infinite-dimensional, but provide an exact description of the marginal process.
(iii) Choose an approximation to obtain finite-dimensional equations.
We will now carry out these steps for our simple example network.

\paragraph*{Description of the marginal process}
Since the first two reactions in Table~\ref{table} do not change the state of $X$, the marginal process consists of two reactions, corresponding to the last two reactions in Table~\ref{table}.
Generally, since the marginal process is no longer Markovian, the transition rates for these reactions will depend on the entire history $x_{[0,t]}$ instead of just on the current state $x(t)$.
However, the transition rate of the fourth reaction in Table~\ref{table} does not depend on the mRNA abundance. Consequently, its marginal transition rate remains unchanged and is given by $c_4 x(t)$. In particular, it depends only on the current state $x(t)$ of the marginal process.
In contrast to this, the rate for the third reaction does depend on the mRNA abundance. As will be derived in Sec.~\ref{sec:marginal_process}, the corresponding marginal transition rate is given by $c_3 \pE[\env{X}(t) \mid x_{[0,t]}]$. This is an intuitive result, expressing the fact that in absence of information about the mRNA abundance, the marginal transition rate is given by the expectation of the transition rate conditional on all available information, i.e. conditional on the entire process history $x_{[0,t]}$.

\begin{figure}
\includegraphics[width=0.48\textwidth]{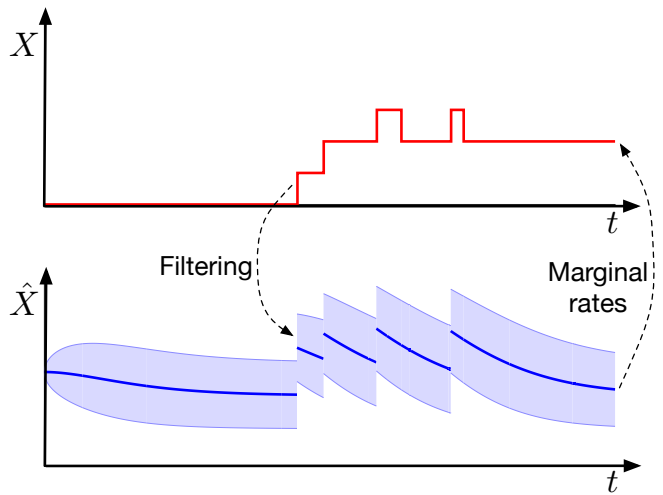}
\caption{(Color online) Schematic illustration of the concepts involved in the construction of the marginal process, based on the example network \eqref{eq:network_geneexp}. Blue curve and shaded area show mean and plus/minus one standard deviation of the filtering distribution. The marginal process trajectory (red curve) drives the evolution of the filtering distribution, and the filtering distribution mean determines the transition rates of the marginal process $X$.
Note that jumps of the filtering distribution occur only if the marginal process increases by a jump. \label{fig:diagram_lhs}}
\end{figure}

\paragraph*{Filtering equation}
We are now tasked with computing the expectation $\pE[\env{X}(t) \mid x_{[0,t]}]$. A convenient way to do this is to derive an evolution equation, driven by the marginal process $X(t)$, for the so-called filtering distribution $\pi_t(\env{x}) := \Pr(\env{X}(t) = \env{x} \mid x_{[0,t]})$ with respect to which this expectation is computed. The resulting equation is called the filtering equation.
As the process $X$ is a jump process, the trajectory $x_{[0,t]}$ is piecewise constant.
The filtering equation for $\pi_t(\env{x})$ will thus consist of two parts: Continuous evolution (described by a differential equation) as long as $X$ remains constant, and discontinuous jumps whenever $X$ jumps. This is schematically illustrated in Fig.~\ref{fig:diagram_lhs}.
As will be derived in Sec.~\ref{sec:reaction_networks}, the continuous evolution is given by
\begin{equation} \label{eq:geneexp_filtering}
\begin{aligned}
\frac{\df}{\df t}\pi_t(\env{x}) =&\; \Omega c_1[\pi_t(\env{x}-1) - \pi_t(\env{x})]\\
&+ c_2[(\env{x}+1)\pi_t(\env{x}+1) - \env{x}\pi_t(\env{x})] \\
&- c_3[\env{x} - \mean{\env{x}}_t]\pi_t(\env{x}).
\end{aligned}
\end{equation}    
Here the expectation $\mean{\env{x}}_t$ is computed with respect to the distribution $\pi_t(\env{x})$ itself.
The first two terms on the right-hand side of \eqref{eq:geneexp_filtering} simply correspond to the ME for the mRNA alone, the dynamics of which do not depend on the protein abundance.
The last term, however, is new and describes how the information that is contained in the trajectory $x_{[0,t]}$ impacts our state of knowledge about the mRNA abundance.
Note that the right-hand side of this equation does not depend on the state of the marginal process $X$. This is because the reaction network has a ``feed-forward'' structure. In general, the filtering equation will depend on the state of $X$. 

As explained above, the filtering distribution will also jump instantaneously whenever the driving process $X$ jumps (see Fig.~\ref{fig:diagram_lhs}). 
At a jump of $X$ at time $t$, the corresponding jump $\pi_{t+} - \pi_{t-}$ of the filtering distribution will depend on which reaction caused the change in $X$.
Since the protein decay reaction does not depend on the mRNA abundance, no information about mRNA abundance is obtained when this reaction fires. Therefore, $\pi_{t+} - \pi_{t-} = 0$ in this case, in analogy to the continuous part \eqref{eq:geneexp_filtering} of the filtering equation in which the protein decay reaction likewise plays no role.
When the protein abundance increases via the third reaction in Table~\ref{table}, however, we instantaneously receive a finite amount of information about the mRNA state. To understand this, note for example that reaction three can fire only if there is at least one mRNA molecule present, i.e. $\env{x} > 0$.
Thus, the filtering distribution immediately after the jump, $\pi_{t+}(\env{x})$, certainly has to satisfy $\pi_{t+}(0) = 0$. As will be shown in Sec.~\ref{sec:reaction_networks}, the jump in the filtering distribution when reaction three fires is given by
\begin{equation} \label{eq:geneexp_filtering_discrete}
\pi_{t+}(\env{x}) = \frac{\env{x}}{\mean{\env{x}}_{t-}} \pi_{t-}(\env{x}).
\end{equation}
In principle, \eqref{eq:geneexp_filtering} and \eqref{eq:geneexp_filtering_discrete} provide a full, exact description of the marginal process, allowing us to compute the marginal transition rates at any time $t$ from the history $x_{[0,t]}$ of the marginal process $X$.
For some simple processes, the corresponding equations can be solved in closed form, as will be demonstrated in Sec.~\ref{sec:simple_example}.
In general, however, these equations constitute an infinite-dimensional system that does not provide a sufficiently simple description of the marginal process dynamics.
We thus have to look for finite-dimensional approximations.

\paragraph*{Finite-dimensional approximation}
Since we are interested only in the expectation $\mean{\env{x}}_t = \pE[\env{X}(t) \mid x_{[0,t]}]$ of the filtering distribution $\pi_t(\env{x})$, it seems reasonable to consider the first-order moment equations for \eqref{eq:geneexp_filtering} and \eqref{eq:geneexp_filtering_discrete}.
However, the equations for the mean $\mean{\env{x}}_t$ are not closed, because the second-order moment $\mean{\env{x}^2}_t$ enters: We obtain
\begin{equation} \label{eq:geneexp_filter_moment}
\frac{\df}{\df t} \mean{\env{x}}_t = \Omega c_1 - c_2 \mean{\env{x}}_t - c_3\left(\mean{\env{x}^2}_t - \mean{\env{x}}_t^2\right)
\end{equation}
from \eqref{eq:geneexp_filtering} and
\begin{equation} \label{eq:geneexp_filter_moment_discrete}
\mean{\env{x}}_{t+} = \frac{\mean{\env{x}^2}_{t-}}{\mean{\env{x}}_{t-}}
\end{equation}
from \eqref{eq:geneexp_filtering_discrete}.
To find a tractable description of the marginal process, we employ moment closure to obtain a finite-dimensional system of equations. As will be explained in more detail in Sec.~\ref{sec:poisson}, in this article we want to obtain the simplest possible description of the (approximate) marginal process, and so choose a first-order closure, incorporating the mean of the filtering distribution only.
A natural choice for such a closure ansatz is the Poisson distribution.
Writing $\theta(t)$ for the mean of the Poisson ansatz distribution, we obtain
\begin{equation} \label{eq:geneexp_filter_moment_poisson}
\frac{\df}{\df t}\theta(t) = \Omega c_1 - c_2 \theta(t) - c_3 \theta(t)
\end{equation}
from \eqref{eq:geneexp_filter_moment} and
\begin{equation} \label{eq:geneexp_filter_moment_discrete_poisson}
\theta(t+) = \theta(t-) + 1
\end{equation}
from \eqref{eq:geneexp_filter_moment_discrete}.
These equations complete our description of the approximate marginal process, which we denote by $X'$ (and refer to as the Poisson-marginal process) to distinguish it from the exact marginal process $X$. Using \eqref{eq:geneexp_filter_moment_poisson} and \eqref{eq:geneexp_filter_moment_discrete_poisson}, we can compute the marginal transition rates at time $t$ based on the full history $x_{[0,t]}$ of the approximate marginal process $X'$.
We use the fact that knowing the history $x_{[0,t]}$ is equivalent to knowing the histories $(y_3)_{[0,t]}$ and $(y_4)_{[0,t]}$ of the two processes $Y_3$ and $Y_4$ (as defined in Sec.~\ref{sec:setting}) that count firings of reactions three and four.
Solving \eqref{eq:geneexp_filter_moment_poisson} and \eqref{eq:geneexp_filter_moment_discrete_poisson} in terms of the process histories, we obtain for the marginal rate of the third reaction the expression
\begin{align*}
c_3 \! \left[e^{-(c_2 + c_3)t} \theta(0) 
+ \! \int_0^t \! e^{-(c_2 + c_3)(t-\tau)} \{\Omega c_1 \df \tau + \df y_3(\tau)\}\right]\!.
\end{align*}
The Stieltjes integral here reduces to a sum, because $(y_3)_{[0,t]}$ is piecewise constant.

\begin{figure*}
\includegraphics[width=0.98\textwidth]{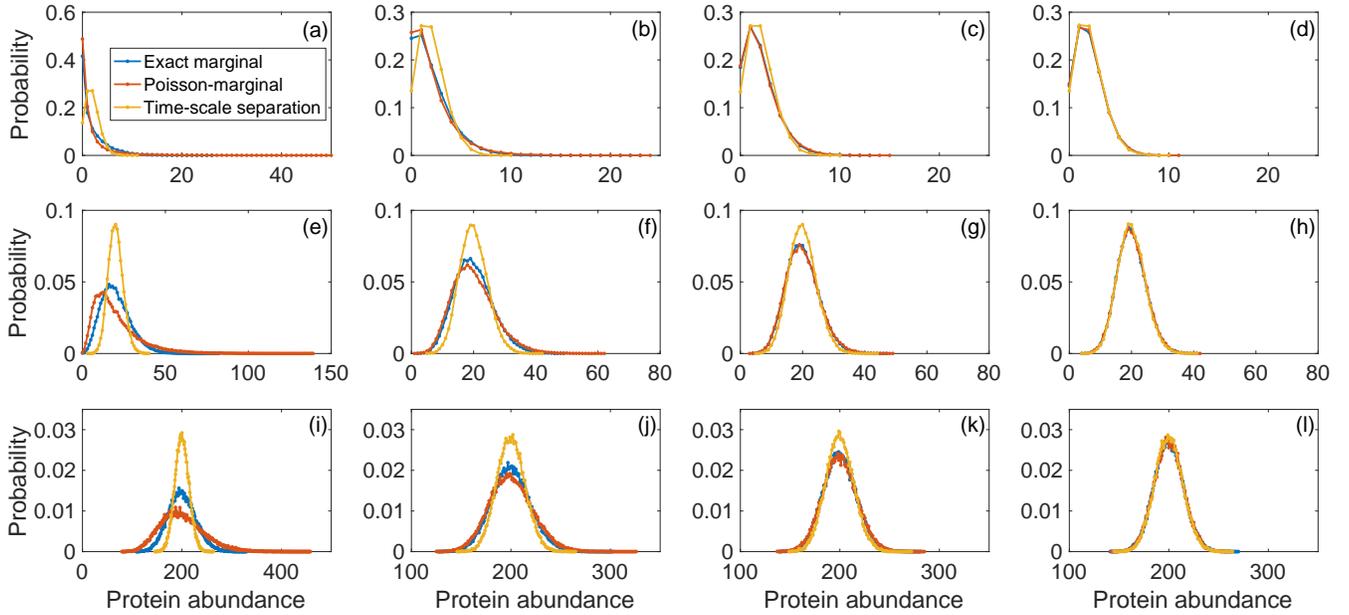}
\caption{(Color online) Monte Carlo evaluation of the approximation quality of the Poisson-marginal process and of time-scale separation. Distributions of protein abundance at stationarity from 50,000 Monte Carlo runs for each case. Parameters were $c_1 = \gamma$, $c_2 = \gamma/2$, $c_3 = 1$ and $c_4 = 0.1$. Rows correspond to system sizes $\Omega = 0.1$ for (a--d), $\Omega = 1$ for (e--h) and $\Omega = 10$ for (i--l). Columns correspond to mRNA process speeds of $\gamma = 0.5$ in (a,e,i), $\gamma = 2$ in (b,f,j), $\gamma = 5$ in (c,g,k) and $\gamma = 25$ in (d,h,l). 
Note that the Poisson-marginal process has a somewhat heavier right tail than the exact marginal process, especially at low system size and low value of $\gamma$.
The algorithm used for stochastic simulation of the marginal process is explained in Appendix~A.
\label{fig:const_geneexp}}
\end{figure*}

In order to evaluate the quality of our chosen approximation, we can compute the mean and the variance of the approximate marginal process $X'$ and of the exact marginal process $X$.
Using results from Sec.~\ref{sec:poisson_closure}, we find that the means of the exact and approximate marginal processes coincide at all times, assuming the initial conditions are chosen appropriately. At stationarity, the means are given by $\mean{x}_{\infty} = \mean{x'}_{\infty} = \Omega c_1 c_3 / c_2 c_4$.
The variances of the processes, however, differ.
We compute the relative error of the variance approximation at stationarity and find
\[
\frac{\mean{x'^2}_{\infty} - \mean{x^2}_{\infty}}{\mean{x^2}_{\infty} - \mean{x}_{\infty}^2} = \frac{c_3^2}{2 c_2 (c_2 + c_3 + c_4)}.
\]
One particular regime where the error vanishes is time-scale separation, when $c_1,c_2 \rightarrow \infty$ with $c_1/c_2$ constant.
It is thus natural to compare our approach with an approximation that directly invokes time-scale separation. As mentioned in the introduction, there exist a large number of approaches. For our simple network, however, there is one particularly natural option (e.g. \cite{cao2005}): We consider the process $\emptyset \overset{k_3}{\longrightarrow} \textrm{Protein} \overset{c_4}{\longrightarrow} \emptyset$ with the rate constant $k_3 = \Omega c_3 c_1 / c_2$.
One easily checks that at stationarity, the time-scale separation ansatz reproduces the correct mean.
We compare the approximation of the full distributions at stationarity numerically in Fig.~\ref{fig:const_geneexp}. We see that the Poisson-marginal process systematically improves on time-scale separation.

\section{The marginal process and the filtering equation} \label{sec:marginal}
In this section, we introduce the marginal process framework in full generality and derive the necessary equations for the case of a general MJP defined on a product-form state-space.
We then specialize to the case of reaction networks, where it is useful to additionally introduce a slightly modified version of the marginal process.

\subsection{The marginal process} \label{sec:marginal_process}
As explained in Sec.~\ref{sec:outline}, the marginal process $X$ is in general no longer Markovian, so that the transition rates at time $t$ will depend on the entire history $x_{[0,t]}$ of the process over the time interval $[0, t]$, instead of just on the current state $x(t)$.
We now proceed to compute these marginal transition rates in a way analogous to \cite{zechner2014a}.

For the marginal process, the probability for a transition into the state $y \in \spS$ to happen in the time interval $[t, t + \Delta t]$, conditional on the process history $x_{[0,t]}$ (and assuming $y \neq x(t)$), is given by
\begin{widetext}
\[
\begin{aligned}
&\pP(X(t + \Delta t) = y \mid x_{[0, t]}) \\
&= \sum_{\env{x}, \env{y}} \pP(X(t+\Delta t) = y, \env{X}(t+\Delta t) = \env{y} \mid X(t) = x(t), \env{X}(t) = \env{x}) 
\pP(X(t) = x(t), \env{X}(t) = \env{x} \mid x_{[0, t]}) \\
&= \sum_{\env{x}, \env{y}}
{L(y, \env{y} \mid x(t), \env{x}) \pP(X(t) = x(t), \env{X}(t) = \env{x} \mid x_{[0, t]})}\Delta t + o(\Delta t)\\
&= \pE[K(y \mid x(t), \env{X}(t)) \mid x_{[0, t]}] \Delta t + o(\Delta t),
\end{aligned}
\]
\end{widetext}
where $K(y \mid x, \env{x}) = \sum_{\env{y}}{L(y, \env{y} \mid x, \env{x})}$ is the total rate for jumps from the state $(x, \env{x})$ leading to any state in $\{y\} \times \spE$.
Thus, the marginal transition rate is given by
\begin{equation} \label{eq:mjp_marginal_hazard}
\pE[K(y \mid x(t), \env{X}(t)) \mid x_{[0, t]}],
\end{equation}
i.e. by the expectation of the total transition rate conditional on the entire history of the marginal process up to time $t$.

The distribution $\pP(\env{X}(t) = \env{x} \mid x_{[0, t]})$ with respect to which the expectation is computed is the filtering distribution for the stochastic process $\env{X}$ given the ``observed'' trajectory $x_{[0, t]}$ of the stochastic process $X$.
The filtering distribution is the solution to the problem of estimating the state of the unobserved variable $\env{X}(t)$ given the available information $x_{[0,t]}$ about the observed variable.

We see that, in order to obtain a useful description of the marginal process, we require a sufficiently simple description of the filtering distribution, or at least of the marginal transition rates $\pE[K(y \mid x(t), \env{X}(t)) \mid x_{[0, t]}]$ that are computed with respect to the filtering distribution.
One way to obtain such a description is to formulate an evolution equation for the filtering distribution driven by the marginal process $X$.
For the case of two fully coupled Markov jump processes, we are not aware of the required results existing in the literature, so we provide an elementary derivation. For an overview of stochastic filtering in general, see \cite{bain2009}.

\subsection{The filtering equation} \label{sec:filtering_equation}
The filtering distribution $\pi_t(\env{x}) := \pP(\env{X}(t) = \env{x} \mid x_{[0,t]})$ is, in principle, defined over the state-space $\spE$ of the nuisance variable.
It is, however, convenient and natural to consider it as a distribution over the joint state-space $\spS \times \spE$ via
\begin{align*}
\pi_t(x, \env{x}) &:= \pP(X(t) = x, \env{X}(t) = \env{x} \mid x_{[0,t]}) \\
&= \delta_{x(t),\, x} \pP(\env{X}(t) = \env{x} \mid x_{[0,t]}),
\end{align*}
where $\delta_{x,\, y}$ is the Kronecker delta. This simply expresses the fact that conditional on $x_{[0,t]}$, the state of $X(t)$ is known to be $x(t)$ with probability one.
Depending on the situation, either of these two views will be more convenient, so that in the following, we will repeatedly switch between considering the filtering distribution to be defined either on $\spE$ or on $\spS \times \spE$.

For the derivations below, the following two operators will be useful:
A summation operator $\opS$ and an evaluation operator $\opP_{y}$ (which depends on a state $y \in \spS$), both of which act on functions $\psi: \spS \times \spE \rightarrow \mathbb{R}$. They are defined by
\begin{equation} \label{eq:op_definition}
\begin{aligned}
{[\opS \psi ]}  &= \sum_{x, \env{x}} \psi(x, \env{x}),\\
[\opP_{y} \psi](x, \env{x}) &= \delta_{y,\, x} \psi(y, \env{x}).
\end{aligned}
\end{equation}

We can now derive the filtering equation. The filtering distribution $\pi_t$ will evolve according to a differential equation in between jumps of the process $X$, and will jump whenever $X$ jumps.
The intuition here is that, over an infinitesimal time-interval $\df t$, if the observed process $X$ does not jump, we receive only an infinitesimal amount of information so that the change in the filtering distribution should also be infinitesimal. When, however, $X$ does jump, we receive a finite amount of information and correspondingly, the filtering distribution has to jump, too. See also Fig.~\ref{fig:diagram_lhs}.

Assuming that we have observed the process $X$ over a time-interval $[0, t + \Delta t]$, these observations can be partitioned into the observations $x_{[0,t]}$ up to time $t$, and the observation $x(t+\Delta t)$. We assume $\Delta t$ sufficiently small such that at most one jump occurred during the time-interval $[t, t+\Delta t]$.
Using Bayes' theorem, we have
\begin{widetext}
\[
\begin{aligned}
&\pP(\env{X}(t+\Delta t) = \env{x} \mid x(t+\Delta t), x_{[0,t]})  \\
&= \sum_{y, \env{y}} \pP(X(t + \Delta t) = x(t+\Delta t), \env{X}(t+\Delta t) = \env{x} \mid X(t) = y, \env{X}(t) = \env{y}) 
\frac{\pP(X(t) = y, \env{X}(t) = \env{y} \mid x_{[0,t]})}{\pP(X(t+\Delta t) = x(t+\Delta t) \mid x_{[0,t]})}  \\
&= \frac{[e^{\Delta t\mathcal{L}} \pi_t](x(t+\Delta t), \env{x})}{\sum_{\env{y}} [e^{\Delta t\mathcal{L}} \pi_t](x(t+\Delta t), \env{y})}
= \frac{\pi_t(x(t+\Delta t), \env{x}) + \Delta t [\opL\pi_t](x(t+\Delta t), \env{x}) + o(\Delta t)}{\sum_{\env{y}} \left\{ \pi_t(x(t+\Delta t), \env{y}) + \Delta t [\opL\pi_t](x(t+\Delta t), \env{y}) \right\} + o(\Delta t)}.
\end{aligned}
\]
\end{widetext}
Multiplying this equation by $\delta_{x(t+\Delta t),\, x}$, using that
\begin{align*}
&\pi_{t+\Delta t}(x, \env{x}) \\
&= \delta_{x(t+\Delta t),\, x} \pP(\env{X}(t+\Delta t) = \env{x} \mid x(t+\Delta t), x_{[0,t]}),
\end{align*}
and noting the definition of $\opP$ and $\opS$ in \eqref{eq:op_definition}, we find
\begin{equation} \label{eq:mjp_filtering_derivation}
\pi_{t+\Delta t}
= \frac{\opP_{x(t+\Delta t)}\pi_t + \Delta t [\opP_{x(t+\Delta t)} \opL \pi_t] + o(\Delta t)}{\delta_{x(t+\Delta t),\, x(t)} + \Delta t [\opS \opP_{x(t+\Delta t)} \opL \pi_t] + o(\Delta t)}.
\end{equation}
In the denominator, we also used that
\[
\sum_{\env{y}} \pi_t(x(t+\Delta t), \env{y}) = \delta_{x(t+\Delta t),\, x(t)}.
\]
We now have to distinguish the cases $x(t+\Delta t) = x(t)$ and $x(t+\Delta t) \neq x(t)$.
When $x(t+\Delta t) = x(t)$, i.e. $X$ remained constant over the time-interval $[t, t+\Delta t]$, we have $\opP_{x(t+\Delta t)}\pi_t = \pi_t$.
Subtracting $\pi_t$ from \eqref{eq:mjp_filtering_derivation}, dividing by $\Delta t$ and taking the limit $\Delta t \rightarrow 0$, we obtain
\begin{equation} \label{eq:mjp_filtering_continuous}
    \ddt \pi_t(x, \env{x}) = [\opP_{x(t)} \opL \pi_t](x, \env{x}) - \pi_t(x, \env{x})[\opS \opP_{x(t)} \opL \pi_t].
\end{equation}
This is the differential equation that the filtering distribution satisfies in between jumps of the process $X$. 
It turns out that \eqref{eq:mjp_filtering_continuous} can also be obtained as an orthogonal projection of the full (joint) ME computed with respect to the Fisher-Rao information metric. This point of view is described in Appendix~B 
and will allow us to better understand the finite-dimensional approximation of the filtering equation introduced in Sec.~\ref{sec:entropic_matching} below.

When $x(t+\Delta t) \neq x(t)$, i.e. when $X$ jumps during the time-interval $[t, t+\Delta t]$, we have $\opP_{x(t+\Delta t)}\pi_t = 0$.
Taking the limit $\Delta t \rightarrow 0$ in \eqref{eq:mjp_filtering_derivation}, we obtain an expression for the filtering distribution immediately after the jump, $\pi_{t+}$, in terms of the filtering distribution immediately before the jump, $\pi_{t-}$, given by
\begin{equation} \label{eq:mjp_filtering_jump}
\pi_{t+}(x, \env{x}) = \frac{[\opP_{x(t+)}\opL \pi_{t-}](x, \env{x})}{[\opS \opP_{x(t+)}\opL \pi_{t-}]},
\end{equation}
where $x(t+)$ is the value of $X$ after the jump.

We now write down expressions \eqref{eq:mjp_filtering_continuous} and \eqref{eq:mjp_filtering_jump} explicitly in terms of the transition rates.
The explicit expressions are simpler if we regard the filtering distribution as being defined only over $\spE$, i.e. $\pi_t = \pi_t(\env{x})$.
We define
\[
R(x, \env{x}) = \sum_{y \neq x}\sum_{\env{y}} L(y, \env{y} \mid x, \env{x}),
\] 
the total rate of those transitions out of state $(x, \env{x})$ that change the $\spS$-component.
In between jumps of $X$, we then have
\begin{equation} \label{eq:mjp_filtering_continuous_explicit}
\begin{aligned}
&\ddt \pi_t(\env{x}) \\
&=\! \sum_{\env{y}} \{ L(x(t), \env{x} \mid\! x(t), \env{y}) \pi_t(\env{y}) - L(x(t), \env{y} \mid\! x(t), \env{x}) \pi_t(\env{x}) \} \\
&\quad - \left\{ R(x(t), \env{x}) - \mean{R(x(t), \env{x})}_t \right\}\pi_t(\env{x}),
\end{aligned}
\end{equation}
where $\mean{R(x(t), \env{x})}_t = \sum_{\env{x}} R(x(t), \env{x}) \pi_t(\env{x})$ denotes the expectation computed using the filtering distribution $\pi_t$.
The first term on the right-hand side of \eqref{eq:mjp_filtering_continuous_explicit} is an ME for the nuisance component $\env{X}$ involving only those transitions which do not change the $\spS$-component of the state.
Note that the corresponding transition rates can still depend on the current state of $X$.
The second term in \eqref{eq:mjp_filtering_continuous_explicit} accounts for the observations.
Here the observations contain information by virtue of the fact that $X$ does not jump as long as \eqref{eq:mjp_filtering_continuous_explicit} is in effect. 
From this equation, we also see that the effect of ``feedback'' from the variable of interest to the nuisance variable is very simple: Because $X$ is constant between its  jumps, $X$ is simply fixed to its current value in the transition rates entering \eqref{eq:mjp_filtering_continuous_explicit}.

When $X$ does jump, so that $x(t+) \neq x(t-)$, the corresponding jump in the filtering distribution is given by
\begin{equation} \label{eq:mjp_filtering_jump_explicit_2}
\pi_{t+}(\env{x}) = \frac{\sum_{\env{y}}L(x(t+), \env{x} \mid x(t-), \env{y}) \pi_t(\env{y}) }{\sum_{\env{x}', \env{y}}L(x(t+), \env{x}' \mid x(t-), \env{y}) \pi_t(\env{y})}.
\end{equation}

The combination of \eqref{eq:mjp_filtering_continuous_explicit} and \eqref{eq:mjp_filtering_jump_explicit_2} with the marginal transition rates \eqref{eq:mjp_marginal_hazard} provides a full description of the marginal process $X$.
For simple processes, these expressions can be evaluated and solved in closed form, as will be demonstrated for a simple reaction network in Sec.~\ref{sec:simple_example}.
Before discussing the example, we specialize the discussion to reaction networks.

\subsection{Reaction networks} \label{sec:reaction_networks}
As we can see from \eqref{eq:mjp_filtering_continuous_explicit}, the transitions of the MJP are naturally partitioned into two groups: Those that change the state of the $\spS$-component, and those that do not.
We will denote by $\notJZ \subseteq \{1, \dots, R\}$ the subset of indices of those reactions which can modify $\vXs$, and by $\JZ = \{1, \dots, R\} \backslash \notJZ$ the indices of all remaining reactions.
This partitioning also results in a partitioning of the counting processes $Y_1, \dots, Y_R$ (defined in Sec.~\ref{sec:setting}) into two processes $\vMs = (Y_j)_{j \in \notJZ}$ and $\vMe = (Y_j)_{j \in \JZ}$ with the former containing the reactions in $\notJZ$ and the latter the remaining reactions in $\JZ$.
Note that the state of the subnet can then be recovered from $\vMs$ alone, while the state of the environment generally requires knowledge of both $\vMs$ and $\vMe$:
\[
X(t) = \sum_{j \in \notJZ}{Y_j(t) \vnu_j},
\quad \env{X}(t) = \sum_{j \in \notJZ}{Y_j(t) \env{\vnu}_j} + \sum_{j \in \JZ}{Y_j(t) \env{\vnu}_j}.
\]

We now specialize the results obtained for the marginal process for general MJPs to the case of reaction networks.
At this point, there arises an issue regarding the precise definition of the history of the marginal process on which we condition in the marginal transition rates \eqref{eq:mjp_marginal_hazard}.
Generally, it can happen that a reaction network contains two different reactions, with different change vectors $(\vnu_i, \env{\vnu}_i)$ and $(\vnu_j, \env{\vnu}_j)$,  for which however the components corresponding to the subnet are identical, $\vnu_i = \vnu_j$.
For example, this is the case for the simple gene expression model with negative feedback
\begin{equation} \label{eq:example_1}
\species{G}_1 \overset{c_1}{\longrightarrow} \species{G}_1 + \species{X}, \quad \species{X} \overset{c_2}{\longrightarrow} \emptyset, \quad \species{G}_1 + \species{X} \underset{c_4}{\overset{c_3}{\rightleftharpoons}} \species{G}_0.
\end{equation}
Here $\species{G}_0, \species{G}_1$ are the two possible states of a gene, and $\species{X}$ is the gene product which is produced when the gene is in state $\species{G}_1$. The gene product can also reversibly bind to the gene and switch it to state $\species{G}_0$, in which production of $\species{X}$ is no longer possible. When the gene product $\species{X}$ is considered to constitute the subnet, the reactions $\species{G}_1 \rightarrow \species{G}_1 + \species{X}$ and $\species{G}_0 \rightarrow \species{G}_1 + \species{X}$ both lead to an increase of $\species{X}$ of size 1.
Similarly, $\species{X} \rightarrow \emptyset$ and $\species{G}_1 + \species{X} \rightarrow \species{G}_0$ both lead to a decrease of $\species{X}$ of size 1.

For such a reaction network, we obtain two different marginal processes depending on whether the history of the process is defined to be just the trajectory $\vxs_{[0,t]}$ (as was done in Sec.~\ref{sec:marginal_process}), or the trajectory $\vms_{[0,t]}$ of the counting processes $Y$ of all reactions which change the subnet.
In the former case we will speak of the marginal process $\vXs$ and in the latter case of the marginal process $\vMs$.
Both marginal processes are meaningful, and only minor changes in the derivations  presented in Sec.~\ref{sec:marginal_process} and \ref{sec:filtering_equation} are necessary.
We will present expressions for both cases, because each version of the marginal process has advantages and disadvantages.

The marginal transition rate for the process $Y$ for reaction $j \in \notJZ$ is given by
\begin{equation} \label{eq:marginal_hazard}
\pE[h_j(\vxs(t), \env{X}(t)) \mid \vms_{[0, t]}]
\end{equation}
where $\vxs(t) = \sum_{j \in \notJZ} y_j(t) \vnu_j$.
This is different from \eqref{eq:mjp_marginal_hazard}, which for reaction networks reads
\[
\sum_{j} \pE[h_j(\vxs(t), \vXe(t)) \mid \vxs_{[0, t]}]
\]
for a transition with change vector $\nu$, and where the summation runs over all $j \in \notJZ$ such that $\vnu_j = \vnu$.

The filtering equation, similarly, exists in two variants, depending on which form of the marginal process we consider.
It turns out however that the continuous part \eqref{eq:mjp_filtering_continuous} of the filtering equation is the same for both variants and explicitly reads
\begin{equation} \label{eq:filtering_continuous_explicit}
\begin{aligned}
&\ddt \pi_t(\vxe) \\
&= \sum_{j \in \JZ}\{h_j(\vxs(t), \vxe-\env{\vnu}_j)\pi_t(\vxe-\env{\vnu}_j) - h_j(\vxs(t), \vxe) \pi_t(\vxe)\}\\
&\quad - \sum_{j \in \notJZ}\left\{h_j(\vxs(t), \vxe) - \mean{h_j(\vxs(t), \env{x})}_{t}\right\}\pi_t(\vxe).
\end{aligned}
\end{equation}
For the marginal process $X$ as defined in Sec.~\ref{sec:marginal_process}, the jump in $\pi_t$ when $X$ jumps is given by
\begin{equation} \label{eq:filtering_jump_explicit_2}
\pi_{t+}(\vxe) = \frac{\sum_{j} h_j(\vxs(t-), \vxe-\env{\vnu}_j)\pi_{t-}(\vxe-\env{\vnu}_j)}{\sum_{j}\mean{h_j(\vxs(t-), \env{x})}_{t-}},
\end{equation}
where the sums in numerator and denominator each run over all reaction indices $j \in \notJZ$ such that $\vnu_j = x(t+) - x(t-)$.
If instead we consider the marginal process $\vMs$, a transition $j \in \notJZ$ leads to a jump in the filtering distribution given by
\begin{equation} \label{eq:filtering_jump_explicit}
\pi_{t+}(\vxe) = \frac{h_j(\vxs(t-), \vxe-\env{\vnu}_j)\pi_{t-}(\vxe-\env{\vnu}_j)}{\mean{h_j(\vxs(t-), \env{x})}_{t-}}.
\end{equation}
The absence of summations in \eqref{eq:filtering_jump_explicit} will be useful in Sec.~\ref{sec:poisson_closure}.
Here we proceed to discuss a simple example for which only the marginal process $X$ is useful.

\subsection{Example: A case with finite-dimensional filtering equations} \label{sec:simple_example}
\begin{figure*}
\includegraphics[width=0.98\textwidth]{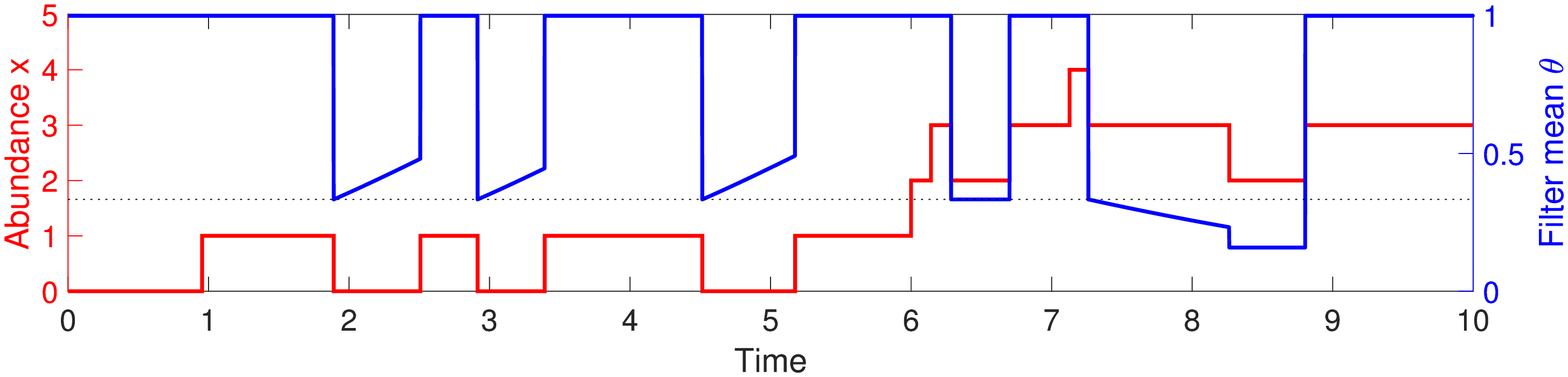}
\caption{(Color online) Example of a sampled trajectory of the process $(X, \Theta)$.
Red curve shows the abundance of gene product $X$ over time.
Blue curve shows the state of the filtering distribution mean $\Theta$.
Note that, as can be seen from \eqref{eq:simple_geneexp_jump}, jumps in $\Theta$  occurring when $\Theta = 1$ always lead to the same value of $\Theta = c_2/(c_2 + c_3)$, indicated by the dotted line. Whether $\Theta$ increases or decreases after the jump depends on the value of $X$. Parameters were $c_1 = 1$, $c_2 = 0.25$, $c_3 = 0.5$ and $c_4 = 2$. Initial state was $X(0) = 0$ and $\Theta(0) = 1$.
\label{fig:simple_geneexp}}
\end{figure*}

We consider the simple gene expression model \eqref{eq:example_1}, with the gene product $\species{X}$ chosen to constitute the subnet.
For this model, every reaction changes the state of $\species{X}$, so that the marginal process $\vMs$ would be equal to the full process and thus of no interest.
Consequently, we instead consider the (one-dimensional) marginal process $X$.
This process has two reactions, $\emptyset \rightarrow \species{X}$ and $\species{X} \rightarrow \emptyset$, with rates at time $t$ given by
\[
c_1 \theta(t) + c_4(1-\theta(t)) \quad \textrm{and} \quad (c_2 + c_3\theta(t))x(t),
\]
respectively, where $\theta(t) = \mean{g_1}_t$ is the filtering distribution mean of the gene state $\species{G}_1$, and where we assumed that only a single copy of the gene is present.
The filtering distribution $\pi_t(g_0, g_1)$, initially defined on $\{0,1\} \times \{0,1\}$, is fully determined by a single number due to the conservation relation $G_0 + G_1 = 1$.
Similarly, for the expectation values with respect to $\pi_t(g_0, g_1)$ we have $\theta(t) = \mean{g_1}_t = \pi_t(0, 1) = 1 - \mean{g_0}_t$.
We can now write down the (one-dimensional) filtering equation using \eqref{eq:filtering_continuous_explicit} and \eqref{eq:filtering_jump_explicit_2}.
In between jumps of $\species{X}$, the result reads
\begin{equation} \label{eq:filtering_continuous_exampel_1}
\ddt \theta(t) = c(x(t)) (1-\theta(t)) \theta(t),
\end{equation}
where $c(x(t)) = c_4 - (c_3 x(t) + c_1)$.
This is solved, for an initial value of $\theta(t_0)$ at time $t_0$, by
\[
\theta(t)
= \frac{\theta(t_0) e^{c(x(t_0))(t-t_0)}}{1 + \theta(t_0) (e^{c(x(t_0))(t-t_0)}-1)},
\]
where we used that $x(t)$ is constant and equal to $x(t_0)$ in between jumps.
When the reaction $\emptyset \rightarrow \species{X}$ fires, the filtering distribution mean $\theta$ jumps to 1.
This is clear because both reactions of \eqref{eq:example_1} that cause a change in $X$ of size $+1$ lead to the gene being in state $\species{G}_1$.
More interesting is the case when the reaction $\species{X} \rightarrow \emptyset$ fires.
Then the jump in the filtering distribution mean is given by
\begin{equation} \label{eq:simple_geneexp_jump}
\theta(t+) = \frac{c_2 \theta(t-)}{c_2 + c_3 \theta(t-)}.
\end{equation}
This completes the description of the marginal process.
If we consider $\theta(t)$ as an auxiliary variable and use it to augment the process state, the resulting process $(X, \Theta)$ is a piecewise-deterministic Markov process with two reactions and deterministic evolution in between jumps given by \eqref{eq:filtering_continuous_exampel_1}.
We show a sample from this augmented process in Fig.~\ref{fig:simple_geneexp}.

While simple systems such as the one discussed in this section can be treated without approximation, more complicated systems will require an approximate solution of the filtering equations, as was already mentioned in Sec.~\ref{sec:outline}.
We address this issue next.

\section{Finite-dimensional approximations of the filtering equation} \label{sec:finite_dim}
The filtering equation is in general infinite-dimensional and, just as the ME, far too complicated for either analytical or numerical solution.
Thus, regardless of whether one is interested in the marginal process for analytical investigation or for stochastic simulation, an approximate treatment of the filtering equation is necessary.

\subsection{Moment equations and moment closure}
A standard approach to obtain a finite-dimensional approximation is moment closure \cite{kuehn2016}, which in the context of stochastic filtering is also known as assumed-density filtering  \cite{brigo1999}.
While moment closures have often been considered \emph{ad-hoc} approximations, a simple variational derivation has been obtained recently \cite{bronstein2017}.

We first derive the filtering moment equations in their general form, starting from \eqref{eq:mjp_filtering_continuous} and \eqref{eq:mjp_filtering_jump}.
The moments of the filtering distribution will evolve according to a differential equation in between jumps of the marginal process, and will jump whenever the marginal process jumps.
Again considering $\pi_t$ to be defined on $\spS \times \spE$, we consider the moment equation for a function $\psi: \spS \times \spE \rightarrow \mathbb{R}$, which can be obtained as demonstrated in \eqref{eq:derivation_moment_eq} and is given by
\begin{equation} \label{eq:filter_moments_continuous_general}
\begin{aligned}
&\frac{\df}{\df t} \mean{\psi}_t  \\
&= \sum_{\vxs, \vxe}{\psi(\vxs, \vxe) \{[\opP_{\vxs(t)} \opL \pi_t](\vxs, \vxe) - \pi_t(\vxs, \vxe)[\opS \opP_{\vxs(t)} \opL \pi_t] \}} \\
&= \mean{\opL^{\dagger} \opP_{\vxs(t)} \left[\psi - \mean{\psi}_t \right]}_t,
\end{aligned}
\end{equation}
in between jumps of $\vXs$, where we used that $\opP_{\vxs(t)}^{\dagger} = \opP_{\vxs(t)}$.
When $X$ jumps, we have
\begin{equation} \label{eq:filter_moments_jump_general}
\mean{\psi}_{t+} = \frac{\mean{\opL^{\dagger} \opP_{\vxs(t+)} \psi}_{t-} }{\mean{\opL^{\dagger} \opP_{\vxs(t+)} \,1}_{t-}}.
\end{equation}
We skip explicit expressions in terms of rates for general MJPs and instead write down the simpler explicit expressions for reaction networks, which read (now for a function $\phi: \spE \rightarrow \mathbb{R}$)
\begin{equation} \label{eq:filter_moment_continuous_explicit}
\begin{aligned}
&\frac{\df}{\df t} \mean{\phi}_t  \\
&= \sum_{j \in \JZ}\left\{\mean{h_j(\vxs(t), \env{x}) \phi(\env{x} + \env{\vnu}_j)}_t - \mean{h_j(\vxs(t), \env{x}) \phi(\env{x})}_t \right\} \\
&\quad - \sum_{j \in \notJZ}\left\{\mean{h_j(\vxs(t), \env{x})\phi(\env{x})}_t - \mean{h_j(\vxs(t), \env{x})}_t \mean{\phi(\env{x})}_t \right\}
\end{aligned}
\end{equation}
in between jumps.
Focusing on the marginal process $Y$, at a jump of $Y$ via reaction $j \in \notJZ$ we have
\begin{equation} \label{eq:filter_moments_jump}
\mean{\phi}_{t+} = \frac{ \mean{h_j(\vxs(t-), \env{x}) \phi(\env{x} +\env{\vnu}_j) }_{t-} }{ \mean{h_j(\vxs(t-), \env{x})}_{t-} }.
\end{equation}

The moment equations are, as is generally the case, not closed: Choosing, for instance, $\phi(\vxe) = \env{x}_n$ for a reaction network to obtain first-order moments, the resulting equations will depend on moments of order higher than one.
In this way, an infinite hierarchy of moment equations is obtained.
We require a way to close the system of equations.

In this paper, we consider a special case of moment closure, which has a dual interpretation based on minimization of relative entropy \cite{ramalho2013, bronstein2017} and on projection using the Fisher-Rao information metric \cite{brigo1999}. 
We next present a derivation analogous to \cite{bronstein2017}, which allows for a unified treatment of the continuous and discrete parts of the filtering equation.

\subsection{Entropic matching} \label{sec:entropic_matching}
A finite-dimensional approximation of a distribution $p(\vxe)$ can be obtained by choosing a distribution from within a finite-dimensional parametric family $p_{\vth}(\vxe)$ with parameters $\theta$.
There are strong arguments \cite{leike2017} for choosing this approximation so that it minimizes the relative entropy 
\[
    D[p \parallel p_{\vth}] = \sum_{\vxe} p(\vxe) \ln \frac{p(\vxe)}{p_{\vth}(\vxe)}.
\]
In the context of the filtering equation, we proceed as follows \cite{bronstein2017}:
Choose a parametric family of probability distributions $p_{\vth}(\vxe)$ depending on  parameters $\vth$ ranging in some open subset of $\mathbb{R}^K$.
Assume that, at time $t$, we have an approximation $p_{\vth(t)}(\vxe)$ of the filtering distribution $\pi_t(\vxe)$ available. As for the filtering distribution itself, we identify the approximation $p_{\vth}(\vxe)$ on $\spE$ with $p_{\vth}(\vxs, \vxe) = \delta_{\vxs(t),\, \vxs}\, p_{\vth}(\vxe)$ on $\spS \times \spE$.

We first consider the continuous part of the filtering equation.
Then a short time-step $\Delta t$ later, $p_{\vth(t)}$ will have evolved to
\[
\begin{aligned}
p(\vxs, \vxe) &= p_{\vth(t)}(\vxs, \vxe) + \Delta t  [\opP_{\vxs(t)} \opL p_{\vth(t)}](\vxs, \vxe) \\
&\quad - \Delta t \, p_{\vth(t)}(\vxs, \vxe)[\opS \opP_{\vxs(t)} \opL p_{\vth(t)}] .
\end{aligned}
\]
We will obtain an approximation to $p(\vxs, \vxe)$ that lies in the parametric family $p_{\vth}$ by choosing parameters $\vth(t+\Delta t)$ to minimize the relative entropy $D[p \parallel p_{\vth(t+\Delta t)}]$.
We then take the limit $\Delta t \rightarrow 0$ to obtain an ordinary differential equation (ODE) for the parameters $\vth$.
Write, for brevity, $\vth = \vth(t)$ and $\tilde{\vth} = \vth(t + \Delta t)$.
Then we have, to first order in $\Delta t$,
\[
\begin{aligned}
&D[p \parallel p_{\tilde{\vth}}] \\
&= \mean{\ln \frac{p_{\vth} + \Delta t \{ \opP_{\vxs(t)} \opL p_{\vth} - p_{\vth}[\opS \opP_{\vxs(t)} \opL p_{\vth}]\}}{p_{\tilde{\vth}}} }_{p} \\
&= \mean{\ln \frac{p_{\vth}}{p_{\tilde{\vth}}}}_{\vth}  \\
& \quad + \Delta t \left[ \mean{ \left\{ \frac{\opP_{\vxs(t)} \opL p_{\vth}}{p_{\vth}}  - [\opS \opP_{\vxs(t)} \opL p_{\vth}]\right\} \ln \frac{p_{\vth}}{p_{\tilde{\vth}}}}_{\vth} + \textrm{const}\right]
\end{aligned}
\]
where ``const'' denotes terms independent of $\tilde{\vth}$, and $\mean{\,\cdot\,}_{\theta}$ denotes an expectation taken with respect to the distribution $p_{\theta}$.
The first term is simply equal to $D[p_{\vth} \parallel p_{\tilde{\vth}}]$, which to second order in $\tilde{\vth} - \vth$ is given by
\[
D[p_{\vth} \parallel p_{\tilde{\vth}}] = \frac{1}{2}(\tilde{\vth} - \vth)^{\dagger} G(\vth) (\tilde{\vth} - \vth),
\]
where $G(\vth)$ is the Fisher information matrix of the parametric family $p_{\vth}$ at parameter value $\vth$, the components of which are given by
\begin{equation} \label{eq:fisher_matrix}
G_{kl}(\vth) = \mean{ \frac{\partial \ln p_{\vth}}{\partial \theta_k} \frac{\partial \ln p_{\vth}}{\partial \theta_l}  }_{\vth}, \quad k,l = 1, \dots, K.
\end{equation}
To minimize $D[p \parallel p_{\tilde{\vth}}]$, we take the derivative with respect to $\tilde{\vth}$ and obtain
\[
\begin{aligned}
0 &= G(\vth)(\tilde{\vth} - \vth)  \\
&\quad - \Delta t \mean{ \left\{ \frac{\opP_{\vxs(t)} \opL p_{\vth}}{p_{\vth}}  - [\opS \opP_{\vxs(t)} \opL p_{\vth}] \right\} \nabla_{\tilde{\vth}} \ln p_{\tilde{\vth}} }_{\theta} \\
&= G(\vth)(\tilde{\vth} - \vth) - \Delta t \mean{ \opL^{\dagger} \opP_{\vxs(t)}^{\dagger} \nabla_{\tilde{\vth}} \ln p_{\tilde{\vth}} }_{\theta} \\
&\quad + \Delta t [\opS \opP_{\vxs(t)} \opL p_{\vth}] \mean{ \nabla_{\tilde{\vth}} \ln p_{\tilde{\vth}} }_{\theta}.
\end{aligned}
\]
Dividing by $\Delta t$, taking the limit $\Delta t \rightarrow 0$ and using that $\mean{\nabla_{\theta} \ln p_{\theta}}_{\theta} = 0$, we get
\begin{equation} \label{eq:entmatch_general}
\begin{aligned}
\ddt \vth &= G(\vth)^{-1} \mean{\opL^{\dagger} \opP_{\vxs(t)} \nabla_{\vth} \ln p_{\vth} 
  }_{\vth}.
\end{aligned}
\end{equation}
This is a closed equation for the parameters $\theta$. Using the resulting approximate solution $p_{\theta}$ of the filtering equation, all necessary expectations, in particular the marginal transition rates, can be computed.

When the process $X$ jumps, the filtering distribution jumps according to \eqref{eq:mjp_filtering_jump}, so that the approximation $p_{\vth(t-)}$ immediately before the jump is updated to
\[
p = \frac{\opP_{\vxs(t+)}\opL p_{\vth(t-)}}{[\opS \opP_{\vxs(t+)}\opL p_{\vth(t-)}]}.
\]
Here too we can obtain a new approximation within the parametric family by minimizing the relative entropy, i.e. choosing $\vth(t+)$ to minimize $D[p \parallel p_{\vth(t+)}]$.
In general, this will be impractical. However, usually one will choose $p_{\vth}$ to be an exponential family
\begin{equation} \label{eq:exp_family}
p_{\vth}(\vxe) = \frac{1}{Z(\vth)}\exp\left\{\sum_{k=1}^K{\theta_k \phi_k(\vxe)}\right\} q(\vxe).
\end{equation}
In this case, minimizing the relative entropy amounts to matching moments, i.e. choosing $\vth(t+)$ so that $\mean{\phi_k}_{\vth(t+)} = \mean{\phi_k}_{p}$ for $k = 1, \dots, K$, which is practical.

The entropic matching equations (applied in the context of filtering for stochastic differential equations) were first proposed in \cite{brigo1999} and derived using a projection argument employing the Fisher-Rao information metric.
This geometrical approach to \eqref{eq:entmatch_general}, which we describe in Appendix~B, is completely analogous to the projection leading to the filtering equation.
In this way, entropic matching is seen to be a very natural way to produce a finite-dimensional approximation to the filtering equation, in addition to the justification provided above.

\subsection{Example: The totally asymmetric exclusion process} \label{sec:tasep}
In this section, we will apply the marginal process framework to the TASEP on the line with open boundaries.
The TASEP \cite{derrida1998} describes particles hopping on $N$ sites $X_1, \dots, X_N$, where each site can be occupied by at most one particle. We take $X_n = 1$ when site $X_n$ is occupied, and $X_n = 0$ otherwise.
If the first site $X_1$ is empty, a particle can enter at a rate $\alpha$.
If site $X_{n+1}$ is empty and site $X_n$ occupied, a particle can move from $X_n$ to $X_{n+1}$ with rate $c$. Finally, a particle at the last site $X_N$ can leave the system with rate $\beta$.

We consider the situation where only the dynamics of the last site $X_N$ is of interest to us, which might serve as a proxy for, say, the flux through the entire system.
Thus, the only transitions which will be retained are the two transitions corresponding to a particle entering or leaving site $X_N$.
For simplicity, in the notation in this section, we do not distinguish between the variable of interest $X_N$ and the remaining variables.
The filtering moment equations for the mean occupancies read
\[
\begin{aligned}
\frac{\df}{\df t} \mean{x_1}_t &= \alpha \mean{1-x_1}_t - c\mean{x_1(1-x_2)}_t, \\
\frac{\df}{\df t} \mean{x_n}_t &= c\mean{x_{n-1}(1-x_n)}_t - c\mean{x_n(1-x_{n+1})}_t, \\
&\quad\; n = 2, \dots, N-2, \\
\frac{\df}{\df t} \mean{x_{N-1}}_t &= c\mean{x_{N-2}(1-x_{N-1})}_t \\
&\quad - c(1-x_N(t)) \mean{x_{N-1}}_t \mean{1-x_{N-1}}_t.
\end{aligned}
\]
As expected and is well known, these contain second-order moments.
Here we are interested in obtaining the simplest possible approximate marginal process.
Thus, we will obtain closed equations in terms of the first-order moments $\mean{x_1}_t, \dots, \mean{x_{N-1}}_t$ only.
A very natural approach to obtain such a closure is to use entropic matching with a product-Bernoulli distribution ansatz:
\[
p_{\vth}(x) = \prod_{n=1}^{N-1}{\theta_n^{x_n} (1-\theta_n)^{1-x_n}}.
\]
We refer to the resulting approximate marginal process as the \emph{Bernoulli-marginal} TASEP.
After application of product-Bernoulli entropic matching, the closed filtering moment equations, in between observations, are given by
\[
\begin{aligned}
\ddt \theta_1(t) &= \alpha (1-\theta_1(t)) - c\theta_1(t)(1-\theta_2(t)), \\
\ddt \theta_n(t) &= c\theta_{n-1}(t)(1-\theta_n(t)) - c\theta_n(t)(1-\theta_{n+1}(t)), \\
&\quad\; n = 2, \dots, N-2, \\
\ddt \theta_{N-1}(t) &= c\theta_{N-2}(t)(1-\theta_{N-1}(t)) \\
&\quad - c(1-x_N(t)) \theta_{N-1}(t)(1-\theta_{N-1}(t)).
\end{aligned}
\]
Unsurprisingly, the resulting equations are identical to a ``naive'' mean-field approximation. 
Note that when $x_N(t) = 1$, i.e. the last site is occupied, the observation term (the last term of the last line) vanishes because a particle cannot enter the last site.

When a particle leaves site $X_N$, no update to the filtering distribution moments is required.
When a particle enters site $X_N$ at time $t$, the update is simply given by
\[
\begin{aligned}
    \theta_n(t+) &= \theta_n(t-), \quad n = 1, \dots, N-2, \\
    \theta_{N-1}(t+) &= 0.
\end{aligned}
\]
This is intuitively clear: Site $X_{N-1}$ is necessarily empty immediately after a particle enters site $X_N$.
The means of the remaining sites are left unchanged because of the product-form closure employed.

\begin{figure*}
\includegraphics[width=0.98\textwidth]{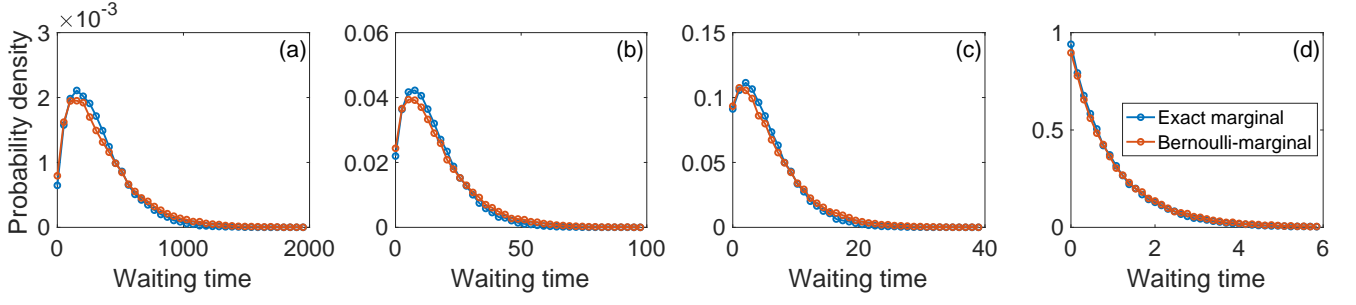}
\caption{(Color online) Numerical evaluation of the accuracy of the Bernoulli-marginal TASEP approximation on $N = 10$ sites at stationarity. Waiting-time distributions for a particle to enter site $X_N$ after the previous particle left $X_N$. Parameters were $\alpha = \beta = 1$ and $c = 0.01$ in (a), $c = 0.1$ in (b), $c = 1$ in (c) and $c = 2$ in (d). Distributions from 100,000 samples.
\label{fig:tasep_1}}
\end{figure*}

We performed Monte Carlo simulations of both the Bernoulli-marginal TASEP and of the full TASEP to compare their behavior.
In Fig.~\ref{fig:tasep_1}, we plot the distribution of waiting times between a particle leaving site $X_N$ and the next particle entering $X_N$ when the process is at stationarity.
The waiting-time distribution varies depending on the parameters of the process.
The Bernoulli-marginal process reproduces the exact results with high accuracy, despite the fact that we have used a very simple closure for the filtering equation.

\section{The product-Poisson marginal process} \label{sec:poisson}
In this section, we will apply our results to general reaction networks.
While the marginal process framework described in Sec.~\ref{sec:marginal} and \ref{sec:finite_dim} is very flexible, it does not provide any indication of how to close the filtering moment equations.
Each choice of closure leads to a different (approximate) marginal process.
In Sec.~\ref{sec:tasep}, we chose what is presumably the simplest-possible non-trivial closure (depending on a single parameter for each variable of the filtering equation) for the TASEP.

In this section, in order to initiate the systematic study of the marginal process framework for reaction networks, we similarly investigate what is arguably the simplest non-trivial closure for reaction networks with mass-action kinetics, one of the most important classes of reaction networks.
Throughout, we focus on the marginal process $Y$ as introduced in Sec.~\ref{sec:reaction_networks}, which is more convenient here.

\subsection{The product-Poisson closure} \label{sec:poisson_closure}
We consider a general reaction network \eqref{eq:reaction_system} with mass-action rates.
In Sec.~\ref{sec:tasep}, we employed product-Bernoulli entropic matching, which lead to  naive mean-field equations.
In the context of general reaction networks, we note that a naive first-order mean-field closure, in which the variance is set to zero, leads
to the vanishing of the second term (corresponding to the observations) in \eqref{eq:filter_moment_continuous_explicit}.
Instead, we will obtain a principled closure by employing entropic matching using a product-Poisson distribution
\[
p_{\vth}(\vxe) = \prod_{n=1}^{\Ne}e^{-\theta_n} \frac{\theta_n^{\env{x}_n}}{\env{x}_n!}.
\] 
Applying product-Poisson entropic matching to a mass-action reaction network leads to equations for the Poisson means that coincide with the macroscopic reaction rate equations. 
This result will then also hold for the first term in \eqref{eq:filter_moment_continuous_explicit} which corresponds to the ``prior'' evolution of the environment species.
In this sense, product-Poisson entropic matching behaves similar (though not identical) to a naive mean-field closure.
However, unlike for a naive first-order closure, the term corresponding to the observations in \eqref{eq:filter_moment_continuous_explicit} does not vanish. We obtain
\begin{equation} \label{eq:filter_poisson_cont}
\ddt \vth = \sum_{j \in \JZ} \Omega c_j f_j(\vxs(t))\hat{g}_j(\vth) \env{\vnu}_{j} 
- \sum_{j \in \notJZ} \Omega c_j f_j(\vxs(t)) \hat{g}_j(\vth) \env{\vs}_{j}
\end{equation}
for the continuous part of the filtering equation, and 
\begin{equation} \label{eq:filter_poisson_disc}
\vth(t+) = \vth(t-) + \env{\vecr}_{j}
\end{equation}
when the marginal process jumps via reaction $j$, where $\env{\vecr}_j = (\env{r}_{1j}, \dots, \env{r}_{\env{N}j})$ and $\env{\vs}_j = (\env{s}_{1j}, \dots, \env{s}_{\env{N}j})$.
Here $\hat{g}_j$ are the mass-action rates for the environment species in their macroscopic form, given by
\[
\hat{g}_j(\vth) = \prod_{n=1}^{\Ne}\left(\frac{\theta_n}{\Omega}\right)^{{\env{s}_{nj}}}.
\]
The simplicity of \eqref{eq:filter_poisson_disc} is the reason for considering the marginal process $Y$.
If we instead consider the marginal process $X$, the corresponding equation is more complicated, and the results obtained in the following would not hold.

In order to better understand the Poisson-marginal process, we can investigate its moment equations. For this purpose, we consider the augmented process $(X', \vTh)$, where $X'$ is the approximate marginal process and $\vTh$ is the stochastic process corresponding to the Poisson means $\theta$. As in Sec.~\ref{sec:simple_example}, $(X', \vTh)$ is a piecewise deterministic Markov process.
For such a process, from the known form of the backwards evolution operator \cite{crudu2012}, 
we obtain the moment equation for a function $\psi(\vxs, \vth)$ in the form
\begin{equation} \label{eq:poisson_marg_moments}
\begin{aligned}
&\frac{\df}{\df t} \mean{\psi}\\
&= \sum_{j \in \JZ}\mean{\Omega c_j f_j(\vxs)\hat{g}_j(\vth) \adj{\env{\vnu}_j} \nabla_{\vth}\psi} \\
&\quad - \sum_{j \in \notJZ} \mean{\Omega c_j f_j(\vxs) \hat{g}_j(\vth) \adj{\env{\vs}_j}  \nabla_{\vth}\psi} \\
&\quad + \sum_{j \in \notJZ} \mean{\Omega c_j f_j(\vxs) \hat{g}_j(\vth)[\psi(\vxs + \vnu_j, \vth + \env{\vecr}_j) - \psi(\vxs, \vth)]}.
\end{aligned}
\end{equation}
Here and in the following, we consider all vectors as column vectors.
In particular (writing for brevity $\chi_j = \chi_j(\vxs, \vth) = \Omega c_j f_j(\vxs) \hat{g}_j(\vth)$), the first-order moment equations are given by
\[
\begin{aligned}
\frac{\df}{\df t} \mean{\vxs} &= \sum_{j \in \notJZ}\mean{\chi_j} \vnu_j = \sum_{j=1}^R\mean{\chi_j} \vnu_j, \\
\frac{\df}{\df t} \mean{\vth} &= \sum_{j \in \JZ}\mean{\chi_j} \env{\vnu}_j - \sum_{j \in \notJZ}\mean{\chi_j} \env{\vs}_j + \sum_{j \in \notJZ}\mean{\chi_j} \env{\vecr}_j \\
&= \sum_{j=1}^R\mean{\chi_j} \env{\vnu}_j,
\end{aligned}
\]
where we used that $\env{\vnu}_j = \env{\vecr}_j - \env{\vs}_j$ and, by definition of $\JZ$, $\vnu_j = 0$ for each $j \in \JZ$.
For a linear reaction network, these equations are identical to the first-order moment equations obtained for the full process $(\vXs, \vXe)$.
Since these equations are closed, we see that the Poisson-marginal process, for a linear reaction network, reproduces the mean of the exact marginal process.

Similarly, we can investigate the relation between the covariance matrices of the Poisson-marginal and the full process by considering the second-order moment equations, which for the augmented Poisson-marginal process $(X', \vTh)$ are given by
\[
\begin{aligned}
\frac{\df}{\df t} \mean{\vxs \adj{\vxs}} &= \sum_{j \in \notJZ} \mean{\chi_j[\vxs \adj{\vnu_{j}} + \vnu_j \adj{\vxs} + \vnu_{j} \adj{\vnu_{j}}]}, \\
\frac{\df}{\df t} \mean{\vxs \adj{\vth}} &= \sum_{j=1}^{R} \mean{\chi_j \vxs \adj{\env{\vnu}_{j}}} + \sum_{j \in \notJZ} \mean{\chi_j[\vnu_{j} \adj{\vth} + \vnu_{j} \adj{\env{\vecr}_{j}}]}, \\
\frac{\df}{\df t} \mean{\vth \adj{\vth}} &= \sum_{j=1}^{R} \mean{\chi_j [\vth \adj{\env{\vnu}_{j}} + \env{\vnu}_{j} \adj{\vth} ]} + \sum_{j \in \notJZ} \mean{\chi_j \env{\vecr}_{j} \adj{\env{\vecr}_{j}}}.
\end{aligned}
\]
Denote by
\[
S = \begin{bmatrix} \mean{\vxs \adj{\vxs}} & \mean{\vxs \adj{\vxe}} \\ \mean{\vxe \adj{\vxs}} & \mean{\vxe \adj{\vxe}} \end{bmatrix}
\quad\textrm{and}\quad 
S' = \begin{bmatrix} \mean{\vxs \adj{\vxs}} & \mean{\vxs \adj{\vth}} \\ \mean{\vth \adj{\vxs}} & \mean{\vth \adj{\vth}} \end{bmatrix}
\]
the matrices of second-order moments for the full process $(X, \env{X})$ and for the augmented Poisson-marginal process $(X', \Theta)$, respectively.
For a linear reaction network, these then evolve according to 
\begin{equation} \label{eq:covar_mat}
\begin{aligned}
\frac{\df}{\df t} S &= A S + S A^{\dagger} + B(t),\\
\frac{\df}{\df t} S' &= A S' + S' A^{\dagger} + B'(t),
\end{aligned}
\end{equation}
with matrices $A, B(t)$ and $B'(t)$.
The difference between the matrices $B(t)$ and $B'(t)$ is given by
\begin{align*}
&B(t) - B'(t) \\
&=\sum_{j=1}^R \mean{\chi_j}_t
\begin{bmatrix}
0 & -\nu_j \adj{\env{s}_j} \\
-\env{s}_j \adj{\nu}_j & \env{\nu}_j \adj{\env{\nu}_j}
\end{bmatrix}
- \sum_{j \in \notJZ} \mean{\chi_j}_t
\begin{bmatrix}
0 & 0 \\
0 & \env{r}_j \adj{\env{r}_j}
\end{bmatrix}.
\end{align*}
Using variation-of-constants to solve \eqref{eq:covar_mat}, we find that the difference between second-order moments of exact and approximate process is given by
\[
S(t) - S'(t) = \int_0^t e^{(t-\tau)A}(B(\tau) - B'(\tau)) e^{(t-\tau)A^{\dagger}} \df \tau,
\]
where we assumed $S(0) = S'(0)$.
In particular, if $B(t) - B'(t)$ is, say, positive semi-definite for all $t \geq 0$, this will also hold for $S(t) - S'(t)$.

\begin{figure*}
\includegraphics[width=0.98\textwidth]{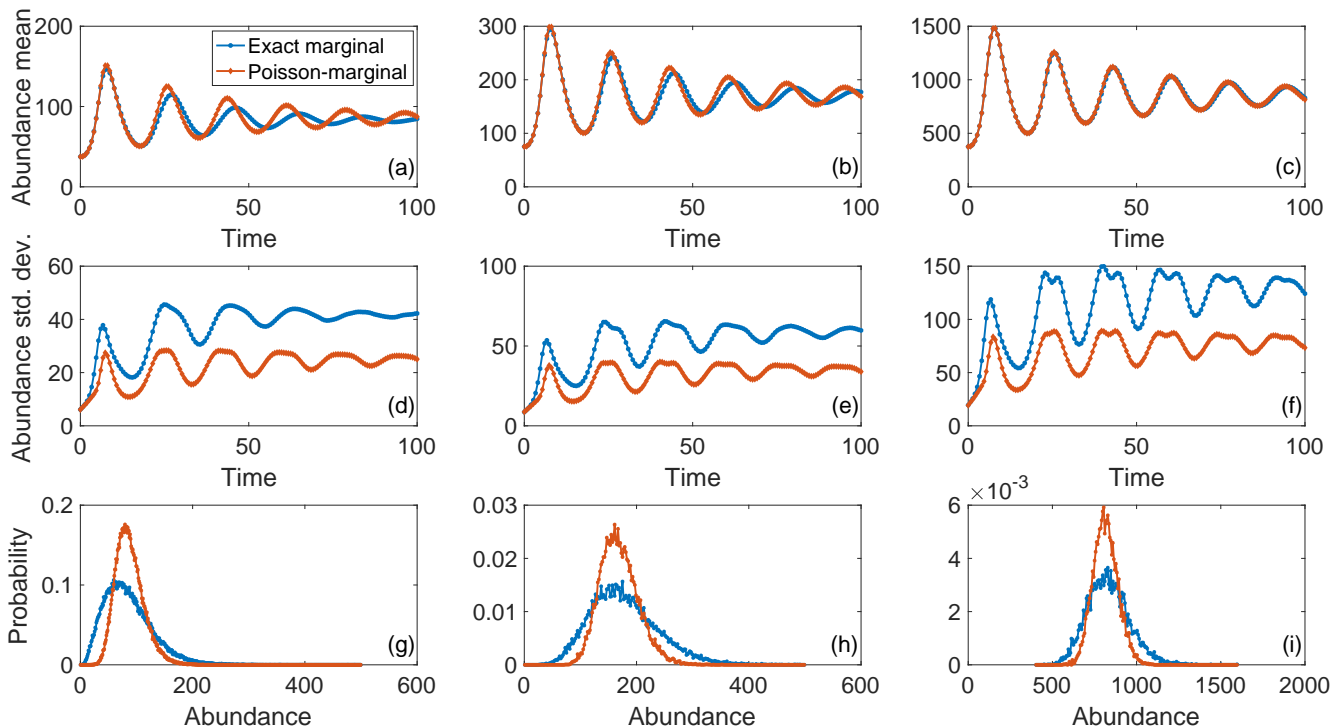}
\caption{(Color online) Monte Carlo evaluation of the accuracy of the Poisson-marginal process for the Lotka-Volterra system at system size $\Omega = 0.5$ in (a,d,g), $\Omega = 1$ in (b,e,h) and $\Omega = 5$ in (c,f,i). 
The mean of the Poisson-marginal process approaches the true mean as the system size increases. In all three cases, the Poisson-marginal process has a smaller standard deviation than the exact process.
(g--i) show the full distributions at time $t = 100$.
Parameters were $c_1 = 1$, $c_2 = 5$, $c_3 = 0.5$, $c_4 = 0.003$ and $c_5 = 0.3$. Initial conditions were of product-Poisson form with means $\mean{\species{X}} = \meanalt{\env{\species{X}}} = 75 \Omega$. The number of simulated trajectories was 100,000 for $\Omega = 0.5$, 20,000 for $\Omega = 1$ and 10,000 for $\Omega = 5$.
\label{fig:lotka}}
\end{figure*}

Even when the reaction network is not linear, the macroscopic rates $\env{g}_j$ coincide with the transition rates $\env{f}_j$ to leading order in the system size $\Omega$ when expressed in terms of concentrations.
One might then expect that in the large system size limit, the mean of the Poisson-marginal process will coincide with the mean of the exact marginal process.
We now investigate these findings numerically on the Lotka-Volterra system
\begin{equation} \label{eq:lotka_system}
\begin{aligned}
\emptyset &\overset{c_1}{\longrightarrow} \env{\species{X}},
&\emptyset &\overset{c_2}{\longrightarrow} \species{X}, \\
\env{\species{X}} &\overset{c_3}{\longrightarrow} 2 \env{\species{X}},
&\env{\species{X}} + \species{X} &\overset{c_4}{\longrightarrow} 2 \species{X},
& \species{X} &\overset{c_5}{\longrightarrow} \emptyset,
\end{aligned}  
\end{equation}
a simple model of predator-prey interaction with oscillatory dynamics.
Here we take the prey species $\env{\species{X}}$ to be part of the environment, while the predator species $\species{X}$ constitutes the subnet.
Numerical results for various system sizes are shown in Fig.~\ref{fig:lotka}.
We see the expected behavior: With increasing system size, the mean of the Poisson-marginal process approaches the exact mean.
We also see that the Poisson-marginal process underestimates the variance of the exact marginal process.

\subsection{Explicit representation of marginal rates}
The representation of the marginal process that we have considered in the previous sections involves auxiliary variables, either the filtering distribution itself or the filtering distribution moments.
It is interesting to represent the (approximate) marginal process in a way that explicitly shows its  memory. 
A subclass of systems for which this is readily done for the Poisson-marginal process are processes with transition rates linear in the environment variables.
Note that this does not imply that the joint reaction network \eqref{eq:reaction_system} is linear. For example, the Lotka-Volterra system \eqref{eq:lotka_system} satisfies this condition.
This will put the marginal process framework in a form more similar to other approaches for obtaining reduced models that have recently been investigated \cite{rubin2014, bravi2016, bravi2017}.

Assume that \eqref{eq:filter_poisson_cont} has the form
\begin{equation} \label{eq:poisson_linear}
    \ddt \vth = F(\vxs(t)) \vth
\end{equation}
for the appropriate matrix $F$, which is explicitly time-dependent through $\vxs(t)$. The extension of the following results to the case where \eqref{eq:poisson_linear} contains an inhomogeneity is obvious.
We write
\[
V(t,\tau) = \overset{\leftarrow}{\mathcal{T}}\exp\left\{ \int_{\tau}^{t} F(\vxs(t')) \df t' \right\}
\]
for the time-ordered exponential (which reduces to a product of finitely many ordinary exponentials because $\vxs(t)$ is piecewise constant).
Noting from \eqref{eq:filter_poisson_disc} that the increments of $\vth$ at jumps of the marginal process are independent of $\vth$, we can represent the solution of the filtering equation as
\[
\vth(t) = V(t,0) \vth(0) + \sum_{i \in \notJZ}{\env{\vecr}_i \int_0^t{V(t,\tau) \df y_i(\tau)}}.
\]
Note that $\vMs$ is a piecewise-constant process with jumps of size $1$, so that the Stieltjes integral reduces to a sum.
Thus, a fully explicit representation for the marginal reaction rate of reaction channel $j \in \notJZ$ at time $t$ of the marginal process is
\[
\mean{h_j}_{\theta} = h_j \left(\! \vxs(t), V(t,0) \vth(0) +\! \sum_{i \in \notJZ}{\env{\vecr}_i \int_0^t{V(t,\tau) \df y_i(\tau)}} \right).
\]

We can apply this result to the Lotka-Volterra system \eqref{eq:lotka_system}.
The equation for the Poisson mean \eqref{eq:filter_poisson_cont} reads (setting $\Omega = 1$ for simplicity)
\[
\ddt \theta = c_1 + c_3 \theta - c_4 x(t) \theta.
\]
The only reaction with a rate depending on $\theta$ is $\env{\species{X}} + \species{X} \overset{c_4}{\longrightarrow} 2 \species{X}$.
Setting $v(t,\tau) = \int_{\tau}^{t}{(c_3 - c_4 x(t'))\df t'}$ and noting that $\env{r}_2 = \env{r}_4 = \env{r}_5 = 0$, we obtain for the marginal reaction rate of this reaction the explicit representation
\[
\mean{h_4}_{\theta} = c_4 x(t) \left[ e^{v(t,0)} \theta(0) + \int_0^t{e^{v(t,\tau)}c_1 \df \tau} \right].
\]

\subsection{Limitations of the product-Poisson closure}
Using a product-Poisson ansatz to close the filtering equation will not always be appropriate.
The most obvious situation where this approach might fail is in the presence of conservation relations among the environment species. 
This will be particularly problematic when there is no intrinsic noise in the environment. A simple example for this behavior would be the gene expression network \eqref{eq:network_geneexp} with input rate $c_1$ and decay rate $c_2$ for the mRNA set to zero.
Irrespective of the initial distribution of mRNA at time zero, the filtering distribution will converge to a unit mass at the true mRNA abundance as the time interval over which the subnet process is observed tends to infinity.
Since for a Poisson distribution the variance is equal to the mean, the vanishing of the variance of the true filtering distribution over time cannot be captured by the Poisson closure.

\section{Conclusion}
In this article, we have illustrated how the marginal process framework, in combination with entropic matching, results in a principled model reduction method for Markov jump processes.
We derived the filtering equation for two fully coupled Markov jump processes in a transparent form that shows its relation to the full master equation, establishing the filtering equation as the result of the application of a projection operator.
The application of a further projection, given by entropic matching, results in a finite-dimensional approximation to the filtering equation and thus in a tractable approximation for the marginal process.
Apart from being a useful tool for the efficient stochastic simulation of the marginal process, it also provides a theoretical understanding of the marginal description.
A particularly simple instantiation of the marginal process framework for mass-action reaction networks, the Poisson-marginal process, was investigated in detail. We derived analytical results for the approximation error for linear reaction networks.
A similar approximation, based on product-Bernoulli entropic matching, was employed for the TASEP.

An interesting question for future investigation is the theoretical analysis of approximations of the filtering equation more accurate than the product-Poisson ansatz or the product-Bernoulli ansatz.
In particular, approximations based on first- and second-order moments might allow one to compare the marginal process framework to other marginalization approaches published previously \cite{rubin2014, bravi2016, bravi2017}.
For this purpose, a variant of the marginal process framework for subnet and environment both modeled by stochastic differential equations would be of interest.

\begin{acknowledgments}
H.K. acknowledges support from the LOEWE research priority program CompuGene.
\end{acknowledgments}

\section*{Appendix A: The marginal simulation algorithm}
Here we describe one possible way to simulate the (approximate) marginal process for reaction networks.
Let
\begin{equation} \label{eq:sim_filtering_cont}
\ddt \vth = v(\vth, x)
\end{equation}
be the differential equation governing the parameters $\vth$ of the (approximate or exact) solution of the filtering equation in between jumps, which in general will depend on the marginal process state $x$. For example, for the Poisson-marginal process, $v$ is given by the right-hand side of \eqref{eq:filter_poisson_cont}.
Similarly, let
\begin{equation} \label{eq:sim_filtering_jump}
\vth_{+} = v_j(\vth_{-}, x_{-})
\end{equation}
be the equation specifying the update to the parameters $\vth$ when the subnet jumps via reaction $j$.
For the Poisson-marginal process, $v_j$ is given by the right-hand side of \eqref{eq:filter_poisson_disc}.

An algorithm \cite{alfonsi2005} based on the modified next reaction method \cite{anderson2007} can be formulated as follows:
The expected reaction rates $\mean{h_j}_{\theta}, j \in \notJZ$ of those reactions that modify the state of $\vMs$ are functions of $\vth$.
We augment the ODE system \eqref{eq:sim_filtering_cont} to include new variables
\begin{equation} \label{eq:simulation_aug}
\ddt \tau_j = -\mean{h_j}_{\vth}, \quad j \in \notJZ.
\end{equation}
The system can then be simulated using Algorithm~\ref{alg}, which samples a trajectory of the (approximate) marginal process over the time-interval $[0,T]$ starting from an initial subnet state $\vxs_0$ and initial parameters $\vth_0$ for the filtering distribution at time 0.
The algorithm has to find the time-point at which a function of the ODE system state crosses a specified threshold (one of the variables $\tau_j$ reaches 0). This is a functionality provided by many ODE solvers, so that the algorithm is straightforward to implement.
\begin{algorithm}[H]
\begin{algorithmic}
\State Set $t \gets 0, \vxs \gets \vxs_0, \vth \gets \vth_0$. \Comment{Initialization}
\For { $j \in \notJZ$ } 
\State Sample $u \sim \textrm{Uniform}(0,1)$.
\State Set $\tau_j \gets -\ln u$.
\EndFor \\

\While {$t < T$} \Comment{Main loop}
\State Solve \eqref{eq:sim_filtering_cont}, \eqref{eq:simulation_aug} until the first variable $\tau_{j_*}$ reaches 0 for

some index $j_{*} \in \notJZ$.
\State Update $\vth \gets v_{j_*}(\vth, x)$.
\State Update $\vxs \gets \vxs + \vnu_{j_*}$.
\State Sample $u \sim \textrm{Uniform}(0,1)$.
\State Set $\tau_{j_*} \gets -\ln u$.
\EndWhile
\end{algorithmic}
\caption{Marginal stochastic simulation algorithm\\
(modified next reaction method)}
\label{alg}
\end{algorithm}

\section*{Appendix B: Filtering equation and entropic matching as projection operations}
\newcommand{\spP}{\ensuremath{\mathbb{P}}}
\newcommand{\spH}{\ensuremath{\mathbb{H}}}
Here we discuss how the continuous part of the filtering equation \eqref{eq:mjp_filtering_continuous} and the entropic matching equation \eqref{eq:entmatch_general} arise as an application of an orthogonal projection (using the Fisher-Rao information metric) applied to the vector field defined by the ME.
Since the results in this paper do not actually depend on any of the results in this appendix, we restrict the discussion to a form which stresses the geometrical significance and neglects any technical difficulties. See \cite{ay2017} for a general treatment of information geometry.

For simplicity, assume that $\spS \times \spE$ is finite, and define the set of probability distributions on $\spS \times \spE$,
\[
\spP = \big \{ p: \spS \times \spE \rightarrow [0,1] \mid \opS p = 1 \big \},
\]
which inherits a manifold structure as a subset of finite-dimensional Euclidean space.
The tangent space at a point $p \in \spP$ is given by
\[
T_p \spP = \{p\} \times \big \{ v: \spS \times \spE \rightarrow \mathbb{R} \mid \opS v = 0 \big \}.
\]
For an MJP $(X,\env{X})$ on $\spS \times \spE$, the master equation $\df p_t(x, \env{x})/\df t = [\opL p_t](x, \env{x})$ defines a vector field on $\mathbb{P}$, the vector attached at a point $p$ being $\opL p$.
We define a basepoint-dependent metric by
\[
g_p(v, w) = \sum_{\substack{x,\env{x}\\ p(x,\env{x}) \neq 0}} \frac{v(x,\env{x}) w(x, \env{x})}{p(x,\env{x})}.
\]
for $v, w \in T_p \spP$.
When restricted to $p \in \spP$ with $p > 0$ everywhere, this is the information metric.
Our extension to other $p$ is somewhat \emph{ad-hoc} but sufficient for our purposes.

It turns out that the continuous part of the filtering equation when $X$ is in state $x(t)$ is obtained simply as an orthogonal projection of the vector field of the full master equation on the tangent space to the submanifold
\[
\mathbb{P}_{x(t)} = \left\{ p \in \mathbb{P} \mid p(x,\env{x}) = \delta_{x(t),\, x}\, \env{p}(\env{x}) \textrm{ for some } \env{p}(\env{x}) \right\}.
\]
From now on, let $p \in \spP_{x(t)}$ with $p(x,\env{x}) = \delta_{x(t),\, x}\, \env{p}(\env{x})$ and assume $\env{p} > 0$ everywhere.
One easily checks that the linear operator
\[
\mathcal{F}_p: T_p \mathbb{P} \rightarrow T_p \mathbb{P}, \quad \mathcal{F}_{p} v = (\opP_{x(t)} - p\opS \opP_{x(t)})v,
\]
satisfies $\mathcal{F}_p^2 = \mathcal{F}_p$ and $g_p(\mathcal{F}_p v, w) = g_p(v, \mathcal{F}_p w)$ for all $v, w \in T_p \spP$, so that $\mathcal{F}_{p}$ is an orthogonal projection.
The projected vector field $p \mapsto (p, \opF_{p} \opL p)$ corresponds to the filtering equation.

The entropic matching equations can similarly be derived as an application of a further projection.
They were in fact first derived in \cite{brigo1999}, using such a geometric approach.
Considering a $K$-dimensional parametric family $p_{\theta}$ of probability distributions on $\spE$, the map $\theta \mapsto \delta_{x(t),\, x} p_{\theta}(\env{x})$ defines a submanifold of $\spP_{x(t)}$ which we denote by $\spP_{x(t)}'$.
From now on, let $p \in \spP_{x(t)}'$ with $p(x,\env{x}) = \delta_{x(t),\, x}\, p_{\theta}(\env{x})$.
The tangent space $T_p \spP_{x(t)}'$ to this submanifold is spanned by the vectors
\[
v_k = v_k(x, \env{x}) = \delta_{x(t),\, x} \frac{\partial p_{\theta}(\env{x})}{\partial \theta_k}, \quad k = 1, \dots, K.
\]
We then find for the Gram matrix
\[
g_p(v_k, v_l) = G_{kl}(\theta),\quad k,l = 1,\dots,K,
\]
i.e. the information metric as defined by \eqref{eq:fisher_matrix}.
The orthogonal projection $\mathcal{Q}_{p}: T_{p} \spP \rightarrow T_{p} \spP$
onto the tangent space $T_p \spP_{x(t)}'$ is given by
\[
\mathcal{Q}_{p}w = \sum_{k,l=1}^K g_p(w, v_k) [G(\theta)^{-1}]_{kl} v_l.
\] 
Because $T_p \spP_{x(t)}'$ is a subspace of $T_p \spP_{x(t)}$, we have $\opQ_{p} \opF_{p} = \opQ_{p}$.
The resulting projected ME, defined by the vector field
\[
p \mapsto (p, \opQ_{p} \opF_{p} \opL p) = (p, \opQ_{p} \opL p),
\]
evolves on the manifold $\spP_{x(t)}'$ when it is started there.
When this is written in terms of the variables $\theta$, we obtain the equations of entropic matching \eqref{eq:entmatch_general}.
Here we also see that entropic matching can be used to directly obtain a finite-dimensional approximation to the filtering equation from the ME, without deriving the filtering equation in an intermediate step.
The derivation presented in Sec.~\ref{sec:entropic_matching} could also be adapted in this way and would then be an application of variational inference \cite{wainwright2008}.

%

\end{document}